\documentclass[twocolumn,showpacs,preprintnumbers,amsmath,amssymb]{revtex4}

\usepackage{amssymb}
\usepackage{mathrsfs}

\usepackage{graphicx}
\usepackage{dcolumn}
\usepackage{bm}
\usepackage{mathrsfs}

\begin{document}


\title{Exclusion limits from data of directional  Dark Matter  detectors 
}

\author{J. Billard}
\email{billard@lpsc.in2p3.fr}
\affiliation{Laboratoire de Physique Subatomique et de Cosmologie, Universit\'e Joseph Fourier Grenoble 1,
  CNRS/IN2P3, Institut Polytechnique de Grenoble, Grenoble, France}

\author{F. Mayet}
\affiliation{Laboratoire de Physique Subatomique et de Cosmologie, Universit\'e Joseph Fourier Grenoble 1,
  CNRS/IN2P3, Institut Polytechnique de Grenoble, Grenoble, France}

\author{D. Santos}
\affiliation{Laboratoire de Physique Subatomique et de Cosmologie, Universit\'e Joseph Fourier Grenoble 1,
  CNRS/IN2P3, Institut Polytechnique de Grenoble, Grenoble, France}

%
%
\date{\today}

\begin{abstract}
Directional detection is a promising search strategy to discover galactic Dark Matter.  
Taking advantage on the rotation of the Solar system around the Galactic center through the Dark Matter halo, it allows to show a direction 
dependence of WIMP events. Even though the goal of directional search is to identify a 
WIMP positive detection,  
exclusion limits are still needed for very low exposure with a rather large background contamination, such as 
the one obtained with prototype experiments. Data of directional detectors are composed  of  energy and a 3D track 
for all  recoiling nuclei. However, to set robust exclusion limits, we focus on the angular part of the event distribution,
 arguing that the energy part of the background distribution is unknown.
 As the angular distributions of both background
and WIMP events are known, a Bayesian approach to set exclusion limits is possible.
In this paper, a new statistical method based on an extended likelihood is proposed, compared to existing ones and is shown to be optimal. 
Eventually, a comprehensive study of the effect of detector configurations  on exclusion limits is presented. It includes the effect 
of having or not sense recognition, a finite angular resolution, taking into account energy threshold as well as  some astrophysical uncertainties.
\end{abstract}

\maketitle

%
%

\section{Introduction}
\label{sec:intro}
In the context of direct detection of non-baryonic Dark Matter (WIMP), an alternative strategy to massive detectors 
\cite{cdms,xenon,edelweiss-armengaud}, 
aiming at high background rejection and planning to scale up to ton-scale, is 
the development of detectors providing an unambiguous positive WIMP signal. 
This can be achieved by searching for a correlation of the WIMP signal 
with  the solar motion around the galactic center, 
observed as a direction dependence of the WIMP flux \cite{spergel}, 
coming from ($\ell_\odot = 90^\circ,  b_\odot =  0^\circ$) in galactic coordinates, which
happens to be roughly in the direction of the Cygnus constellation. This is generally referred to as directional  detection of Dark Matter and several projects of detector are being developed for 
this goal \cite{MIMAC,Drift,mit,newage,white}.\\
The major advantage of directional detection is the fact that the angular distribution of WIMP events  is pointing toward the Cygnus constellation while  
  the background one is isotropic. This opens the  possibility to really identify the WIMP signal as
such. The Discovery parameter in this case is a signal pointing toward Cygnus as emphasized in \cite{billarddisco,greendisco} since it cannot
be mimicked by background. Recently, a statistical map-based analysis has been developed 
\cite{billarddisco}, showing the possibility to extract from  data samples of 
forthcoming directional detectors,  both the main direction of the incoming events, thus proving the
galactic origin of the signal,  and the number of WIMP events contained in the map thus constraining the WIMP-nucleon cross section.\\
Even though the goal of directional experiments is to identify a WIMP positive detection, 
exclusion limit is still needed, for very low exposure with a rather large background contamination, such as 
the one obtained with prototype directional experiments.\\
The directional insensitive direct detection strategy is based on the measurement of the recoil energy ($dN/dE_R$) and a dedicated statistical method has 
been developed \cite{yellin1} to optimize the exclusion limit obtained from a given set of data. In addition to the recoil
energy, directional detectors also provide, for each track,  the scattering angle ($\Omega$). Recently a 2D extension of the 
previous method has been proposed \cite{henderson,yellin2}. It is based on the double-differential spectrum ($d^2N/dE_Rd\Omega$) and
allows to account for all information given by a directional detector. As a matter of fact, the energy spectrum of the background is 
unknown \cite{green.masse,bernal.masse}, while its angular spectrum is expected to be isotropic in the galactic rest frame. 
We take advantage on this
point by proposing a new method, taking into account only the angular part of the spectrum  ($dN/d\Omega$) 
in a given recoil energy range. As no assumption on the background energy dependence is needed, 
robust and conservative exclusion limits may be provided. We also argue that upcoming directional data may potentially be contaminated with a rather large
amount of background events and this method is intended  to cope with low signal to noise ratio, when identification is not possible \cite{billarddisco}. 
This method is based on a likelihood analysis of events that are supposed to be composed of both 
 background and WIMP events.\\
The paper is organized as follows. 
In section \ref{sec:detection} we will first  introduce the directional detection framework. In section \ref{sec:methods}, 
we will introduce the  different statistical methods used to set WIMP-nucleon cross section limits : the existing one 
({\it Poisson}, {\it Maximum Gap} 
 \cite{yellin1}) and a  new  one, referred to as the {\it Directional Likelihood} method. They will be compared in section \ref{sec:Comparison} for various detector configurations such as ideal detector, 
 with or without sense recognition,  finite angular resolution, and threshold effect. It allows to evaluate the impact of experimental uncertainties 
 on each method. The effect of 
 some astrophysical uncertainties like the WIMP local velocity dispersion and the local dark matter density $\rho_0$ will also be discussed.

\section{Directional detection: experimental and theoretical framework}
\label{sec:detection}
Several directional detectors are being developed and/or operated : 
MIMAC~\cite{MIMAC}, DRIFT~\cite{Drift}, DM-TPC~\cite{mit} and NEWAGE~\cite{newage}. A detailed overview of the status of experimental efforts devoted to directional
Dark Matter detection is presented in \cite{white}. Directional detection of Dark Matter requires track reconstruction  of
 recoiling nuclei down 
to a few keV. This can be achieved with low pressure gaseous detectors \cite{sciolla} and several gases have been suggested : 
$\rm  CF_4,^{3}He+C_4H_{10}$ or  $\rm CS_2$. 
Both the energy and the track of the recoiling nucleus need to be precisely measured. Ideally, recoiling tracks should be 3D 
reconstructed as the required exposure to reject the background hypothesis is decreased by an order of magnitude between  the 2D readout and 3D readout \cite{green1}. 
Sense recognition of the recoil track ({\it head-tail}) is also a key issue for
directional detection \cite{dujmic,burgos,majewski} and its impact on exclusion limits calculation will be discussed here.
The study is done for an ideal, but realistic, detector which could be within reach in a few years.
We consider a 10 kg $\rm CF_4$ detector, operated at 50 mbar and allowing 3D reconstruction of recoiling tracks.
 A recoil energy range [$E_{R_1},E_{R_2}$] is chosen between 5 and 50 keV. 
The lower bound of the energy range is due to the threshold ionization energy taking into account the
 quenching factor. As most of the WIMP events are concentrated at low recoil energy, an upper bound is
 chosen to limit the background contamination of the data. Indeed, for a WIMP mass of 100 GeV.c$^{-2}$, 70\% of the recoils are between 5 keV and 50 keV and only 10\% are above
  50 keV  in the case of an escape velocity taken as $v_{esc}=\infty$ and a form factor $F(E_R)$ approximated to one
 \footnote{Using the Born approximation, in the case of a spin-dependent interaction, the
form factor is given by the Fourier transform of a thin shell \cite{lewin} leading to: $F^2({\rm 5~keV}) = 0.99$ and $F^2({\rm 50~keV}) =
0.9$ in the case of a $\rm ^{19}F$ target, justifying our approximation. Note that in the case of heavier targets, this approximation would be no longer valid.}. 
 Thus, increasing the upper bound would lead to a potentially weaker signal to noise ratio.
In section \ref{sec:Comparison}, we will study the effect of having or not sense recognition,
 degrading the  angular resolution as well as varying the energy threshold.\\ 
Through this study, we  focus on the simplest model for the Milky Way halo :  the isotropic isothermal sphere  
in which the WIMP velocity follows
 a Maxwellian distribution defined in the laboratory rest frame as 
\begin{equation}
f(\vec{v}) = \frac{1}{(2\pi\sigma^2_v)^{3/2}}\exp\left (-\frac{(\vec{v} + \vec{v}_{\odot})^2}{2\sigma_v^2}\right )
\label{eq:HaloModel}
\end{equation}
with a dispersion $\sigma_v = v_0/\sqrt{2}$ where $v_0 = 220 $ km.s$^{-1}$
  is the circular speed at large radii. Effect of uncertainties on $v_0$, {\it i.e} on the local WIMP velocity dipersion, on  exclusion limits 
  will be discussed in sec.~\ref{sec:vo}. 
    We consider a detector velocity equal to the 
  tangential component of the Sun's motion around the Galactic center
$v_{\odot} = 220 \pm 20$ km.s$^{-1}$, neglecting the Sun's peculiar velocity and the Earth's orbital velocity about the Sun. As a matter of fact, 
their contribution  to the detector velocity 
is smaller than the uncertainty on $v_{\odot}$  \cite{morgan1}.
Using the galactic coordinates  $(\ell, b)$, the WIMP  velocity is written in the galactic rest frame as~:
$$\vec{v} = v\ (\cos \ell \cos b \ \hat{x}  + \sin \ell \cos b \ \hat{y} + \sin b \ \hat{z})$$
where $\hat{x}$ points toward the galactic center, $\hat{y}$ in the direction of the solar motion and  $\hat{z}$ toward the galactic
 north pole.  
The WIMP-induced recoil events are then computed by generating random incident WIMP velocities from $vf(\vec{v})$ and assuming 
an isotropic elastic scattering in the center of mass frame. Then, the recoil energy ($E_R$) is given, in the laboratory frame, by :
\begin{equation}
E_R = \frac{2m^2_{\chi}m_Nv^2}{(m_{\chi} + m_N)^2}\cos^2 \theta_R
\label{RecoilEnergy}
\end{equation}
with  $m_{\chi}$ the WIMP mass, $m_N$ the mass of the target and $\theta_R$ the recoil angle with respect to the WIMP direction.\\
Thus, this Monte Carlo WIMP event generation can be used both for simulating experiments and for evaluating theoretical angular
distributions. Then, in the case of the binned version of the {\it Directional Likelihood method}, described in sec. \ref{DLM}, the theoretical event distributions of
WIMP events $S$ and
background events $B$ are estimated using 10$^8$ Monte Carlo generated events. In the case of the background, the events are randomly generated according to an
 isotropic distribution.
However, we have also developed an unbinned definition of our new {\it Directional Likelihood method} (see sec. \ref{DLM}). In this case, the $S$ and
 $B$ distributions are analytically calculated. Indeed, as the spherical isothermal halo model leads to a WIMP-induced recoil distribution which presents a
 spatial distribution of events with an axial symmetry along the
$\hat{y}$ axis,  we will use the $\cos\gamma$ distribution of the events with $\cos\gamma = \hat{y}.\hat{r}$, 
$\hat{r}$ being the recoil direction in the galactic coordinates. 
In order to evaluate $S(\cos\gamma)$ (see sec. \ref{DLM}), we used the following theoretical distribution \cite{spergel}:
\begin{equation}
\frac{dR}{d\cos\gamma} = \kappa\int_{E_{R_1}}^{E_{R_2}}\exp\left[ -\frac{\left( v_{\odot}\cos\gamma - v_{min}\right)^2}{2\sigma^2_v} \right]dE_R
\label{dRdCosGamma}
\end{equation}
$\kappa$ is a normalization factor, 
$v_{min} = \sqrt{m_N E_R/2m_r^2}$ corresponds to the minimal WIMP velocity  required to produce a nuclear recoil of energy $E_R$ and $m_r$ is the WIMP-nucleus reduced mass.
Moreover, it is worth noticing that due to this axial symmetry, there is no loss
of information when looking at 1D angular distribution ($\cos\gamma$) rather than 2D angular spectrum ($\ell$, $b$). The theoretical background distribution $B(\cos\gamma)$
corresponds to a flat distribution as the background is taken to be isotropic.\\

The expected number of WIMP 
events $\mu$ is calculated by integrating the theoretical recoil energy distribution over the experimental energy range $E_{R_1}$ to $E_{R_2}$:
\begin{equation}
\mu(\sigma) = \frac{\sigma\rho_0\xi}{2m^2_rm_{\chi}}\int_{E_{R_1}}^{E_{R_2}}F^2(E_R)\int_{v_{min}}^{\infty}\frac{f(\vec{v})}{v}d^3v,
\label{eq:events}
\end{equation}
with $\sigma$ the WIMP-nucleus elastic scattering cross section, $\rho_0$
 the local dark matter density and $\xi$ the exposure. In the following we define $\sigma_0$ the WIMP-nucleon cross section directly related to $\sigma$ 
 in the pure-proton approximation, see sec. \ref{prospects}.


\section{WIMP-nucleon cross section limits calculation methods}
\label{sec:methods}

In this section, we present the existing statistical methods, {\it Poisson} and {\it Maximum Gap} \cite{yellin1}, and we propose a  
new  one dedicated to directional detection, referred to as {\it Directional Likelihood} method.

\subsection{Poisson method}
\label{sec:Poisson}
The simplest way to derive a cross section upper limit from a given experimental dataset, is to use the Poisson method. This is the most
conservative approach because it assumes no knowledge on neither the background nor the WIMP event distribution. Each recorded event is interpreted 
as a WIMP one. Hence, the aim of this method is to define the value of $\sigma_0$ corresponding to a 
number of signal events, $\mu_{exc}$, which is excluded at a given confidence level (CL) according to the Poisson distribution. 
The confidence level is then given by $1 - \alpha(\mu)$, with
\begin{equation}
\alpha(\mu) = e^{-\mu}\sum_{m = 0}^{N}\frac{\mu^m}{m!}
\end{equation}
and N is the total number of events contained in a dataset. 
Then, the excluded value of $\sigma_0$ at 90\% CL is deduced from the value of $\mu_{\rm exc}$, which satisfies $1 - \alpha(\mu_{exc}) = 0.9$, using equation
(\ref{eq:events}).\\ 
As in the Poisson method  all events are considered as a WIMP signal, any background observed events will lead to overly estimated  
upper limits. Even though it has been shown \cite{yellin1,henderson} that, in the case of a pure WIMP signal, the Poisson 
method gives more restrictive upper limits than the Maximum Gap method, in this paper we are only interested in a dataset 
without WIMP events or with a large background contamination, which seems to be closer to real data from upcoming 
directional detectors \cite{white}. We argue that in the case of a pure WIMP signal, directional detection should lead to an identification
of a positive WIMP detection, using the methods presented in \cite{billarddisco,greendisco}.
 
\subsection{1D Maximum gap method}
In order to deal with an unknown background contamination, which is the case with  direction-insensitive direct detection ({\it i.e.} 
measuring only the energy of the recoil), two statistical methods have been proposed by S. Yellin \cite{yellin1} : "the maximum gap" and
"the optimal interval". In this paper, we define a modified version of the maximum gap method applied to 1D directional 
data, {\it i.e.} the angular part of the spectrum ($dR/d\cos\gamma$ distribution, see sec. \ref{sec:detection}). 
 Then, for a given value of $\sigma_0$ and a given dataset of N recorded events, we obtain a set of (N+1) gaps $x_i$ which are calculated as \cite{yellin1}:
\begin{equation}
x_i = \int_{\cos\gamma_i}^{\cos\gamma_{i+1}} \frac{dR}{d\cos\gamma}(\sigma_0)d\cos\gamma
\end{equation}
with $\cos\gamma_0 = -1$ and $\cos\gamma_{N+1} = 1$ which are the lower and upper bounds of the $dN/d\cos\gamma$ distribution. The maximum
gap $x$ is then defined as the largest $x_i$ obtained for a given experiment. The interest of the "maximum gap" method is the fact that the distribution of $x$ is
 independent of the theoretical
distribution of the events. It depends only on $\mu$ and not on the shape of the event distribution. We refer the reader to \cite{yellin1} for a  detailed discussion on
setting limits using the Maximum Gap method. 

 
We do not use full information from the datasets, {\it i.e.} $d^2R/dE_Rd\cos\gamma$, as in \cite{henderson}, since 
the performance of the method strongly depends on the assumed background distributions. As the energy 
distribution of background events is unknown, we prefer to consider 
only the angular part of the distribution  for which the background is known to be isotropically distributed,  leaving aside the energy information of recorded 
events.

\subsection{Directional Likelihood methods}
As stated above, we restrict the information to the angular distribution of the events. In this case,
the background is well understood and then, we have a theoretical distribution for both WIMP events and
 background ones thus allowing a Bayesian calculation of exclusion limits using a likelihood analysis. 
 
\subsubsection{Definition}
\label{DLM}

The {\it Directional likelihood method} is based on a recent paper \cite{billarddisco} which aims at distinguishing a genuine WIMP signal in a
directional dataset. This likelihood definition allows to recover the main direction of the recoils and the number of WIMP events
contained in the recoil map in order to identify a detection of particles from the galactic Dark Matter halo and to evaluate its significance. It has been shown
that this analysis tool gives satisfactory results on a large range of exposure and background contamination levels. But, for very low number of
WIMP events and for very large background fractions, when this method failed at recognizing a WIMP detection, obviously 
 an exclusion limit should be derived.  This will be the case for the very first results of directional
detectors with low exposures.\\

The total number of recorded events N is considered as the sum 
of $n_s$ signal events and $n_b$ background events, {\it i.e.} $N = n_s + n_b$,
 where both $n_s$ and $n_b$ can be regarded as Poisson variables with means $\mu_s$ and $\mu_b$ respectively.
The fact that both signal and background angular spectra are well known allows to derive upper limits using the Bayes' theorem. In the case of flat priors  
for both $\mu_s$ and $\mu_b$, and taking the evidence as  a normalization factor, it is reduced to 
$$
P(\mu_s,\mu_b|\vec{D}) \propto \mathscr{L}(\mu_s,\mu_b)
$$
 where $\vec{D}$ refers to the characteristics of the data, as the total number of recorded events $N$, their direction and
their energy. 

However, in order to incorporate in the likelihood definition both the information contained in the measured angular spectrum and the fact that $N$ is a Poisson variable
 of mean $\mu_N = \mu_s + \mu_b$, an extended likelihood function is used: 
\begin{equation}
\mathscr{L}(\mu_s,\mu_b) = \frac{(\mu_s + \mu_b)^N}{N!}e^{-(\mu_s + \mu_b)}\times L(\mu_s,\mu_b)
\end{equation}
In the following,  two different definitions of  $L(\mu_s,\mu_b)$ are proposed, in the case of binned or unbinned data.\\

When using binned data, the following definition of $L(\mu_s,\mu_b)$ is used :
$$
  L(\mu_s,\mu_b) = \prod_{i=1}^{N_{\rm pixels}} P\left( \frac{\mu_s}{\mu_s + \mu_b} S_i + \frac{\mu_b}{\mu_s + \mu_b}B_i|M_i\right)
$$
where $M$ corresponds to the observed recoil map, $S$ and $B$ refer to the signal (WIMP) and background theoretical distributions (see sec. \ref{sec:detection}). The
probability $P$ is then evaluated at each bin with a poissonian distribution. Here the bins are the pixels of the observed recoil map
represented in galactic coordinates. An angular resolution of 15$^{\circ}$ (FWHM) is chosen \cite{santos}, 
corresponding to a number of equal area pixels of $N_{\rm pixels} = 768$.\\

Unbinned data are often preferred and we propose a unbinned version of the definition of $L(\mu_s,\mu_b)$, defined as~:
$$
L(\mu_s,\mu_b) = \prod_{i=1}^{N} \left(  \frac{\mu_s}{\mu_s + \mu_b} S(\cos\gamma_i) + \frac{\mu_b}{\mu_s + \mu_b}B(\cos\gamma_i)\right)
$$
with $S$ and $B$ the theoretical distributions, $dR/d\cos\gamma$ (see sec. \ref{sec:detection}), of WIMP and Background events respectively.

\subsubsection{Setting exclusion limits using $P(\mu_s,\mu_b|\vec{D})$}

\begin{figure}[t]
\begin{center}
\includegraphics[scale=0.35,angle=270]{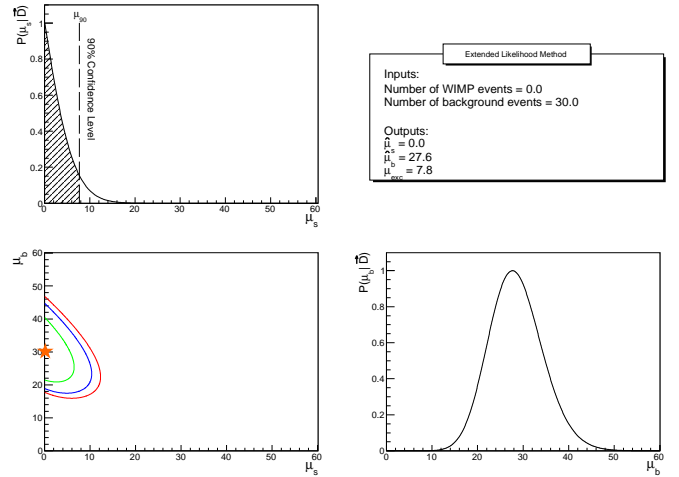}
\caption{Representation of $\mathscr{L}(\mu_s,\mu_b)$ in the case of 0 expected WIMP events and 30 expected background events.
 The lower left panel presents the 2 dimensional likelihood distribution with the 68\%, 90\% and 95\% CL contours. The orange star represents the input values of $\mu_s$
 and $\mu_b$.
  The left upper panel and the right lower panel represents the marginalized distributions $P(\mu_s|\vec{D})$ and $P(\mu_b|\vec{D})$, respectively.}  
\label{fig:Likelihood2D}
\end{center}
\end{figure}

Using a Bayesian approach, the probability density function of the parameter of interest $\mu_s$ can be derived by marginalizing $P(\mu_s,\mu_b|\vec{D})$ over the parameter
$\mu_b$. 
Then, the value of $\mu_{\rm exc}$ is obtained by solving:
\begin{equation}
\int_0^{\mu_{\rm exc}} P(\mu_s|\vec{D}) \ d\mu_s = 0.9,
\end{equation}
Thus, the value of the excluded cross section at 90\% CL is deduced from $\mu_{\rm exc}$ using equation (\ref{eq:events}).
As an illustration of the method, figure \ref{fig:Likelihood2D} presents  the results from the
calculation of $\mathscr{L}(\mu_s,\mu_b)$ in a working example with 0 expected WIMP event and 30 expected background events. 
The probability density functions of
$\mu_s$ and $\mu_b$ are derived from the marginalization of the 2 dimensional likelihood over the $\mu_b$ and $\mu_s$ parameters respectively. The calculation of
$\mu_{\rm exc}$ is also illustrated and represented on the left upper panel of figure \ref{fig:Likelihood2D}. Hence, on this example of 30 simulated events, the method allows
to exclude at 90\% CL a number of WIMPs greater than 7.8.
At last, a strong correlation ($\rho \sim -1$) between the parameters $\mu_s$ and $\mu_b$ is observed on the 2 dimensional representation of
 $\mathscr{L}(\mu_s,\mu_b)$ and is explained by the fact that $n_s = N - n_b$. 
 
\subsubsection{Comparison between binned and unbinned likelihood definitions}
When using binned data, bias can be introduced in the analysis and thus, whenever possible, unbinned data analysis are preferred.
For each configuration and each result presented in this paper, both binned and the unbinned analyses have been done, using the proper 
likelihood definition. No difference between them has been seen, due to the fact that, in our case,
the binning (pixels area) is thin enough and the theoretical distributions are smooth and much wider than the pixels. 
As the effect of bining is null, using binned data do not introduce any bias in the calculation of $P(\mu_s,\mu_b|\vec{D})$ and then in the estimation of the excluded cross
section. 
Hence,  only the results obtained with the binned likelihood definition,  
referred to as the {\it Directional Likelihood method}, is presented hereafter.
 
\subsection{Discussion on limit setting methods}
\label{sec:quescution}
\begin{figure}[t]
\begin{center}
\includegraphics[scale=0.35,angle=270]{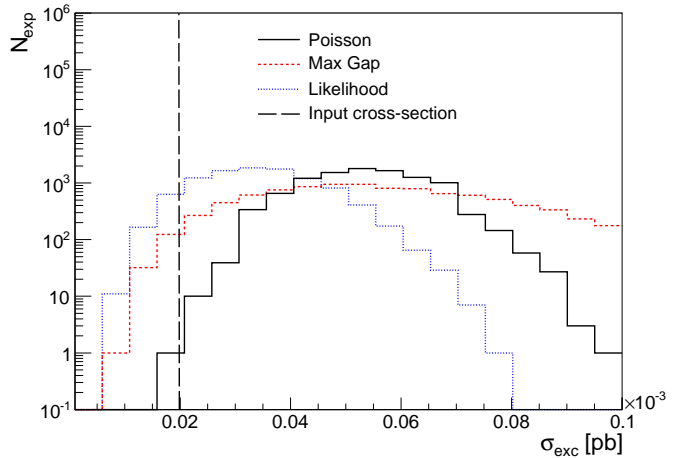}
\caption{Frequency distributions of the excluded cross section $\sigma_{\rm exc}$ at 90\% CL for each method with 10 expected WIMP events and 10 expected background events. The
Poisson, Maximum Gap and Likelihood methods are represented by the black solid line, the red dashed line and the blue dotted line respectively. The coverage for each methods
are (99.99 $\pm$ 1)\%, (98.76 $\pm$ 1)\% and (93.39 $\pm$ 1)\% respectively.}  
\label{fig:Frequency}
\end{center}
\end{figure}

In order to evaluate the frequency distributions of the excluded cross section at 90\% CL, see figure \ref{fig:Frequency}, we used 10,000
toy Monte Carlo   experiments
 for each case and for the three statistical methods.

From each distribution, we can derive the median value of the excluded WIMP-nucleon cross section $\sigma_{med}$ and the RMS given by the Poisson, Maximum Gap and the
Likelihood methods. As an example, on figure \ref{fig:Frequency} we consider a case with 10 expected WIMP events and 
10 expected background events. The coverage is defined as the fraction of experiments leading to  an excluded 
cross section above the input value. The values deduced from the study  
 are  : (99.99 $\pm$ 1)\%, (98.76 $\pm$ 1)\% and (93.39 $\pm$ 1)\% for the Poisson, Maximum
Gap and likelihood methods respectively. The fact that the coverage is greater than 90\% for the three different methods means that in the case of a background contamination,
the excluded cross section is overly estimated. In the case of the Poisson method, this over estimation comes from the fact that each recorded event is considered
 as a WIMP event. Contrary to the Poisson method, the Likelihood and the Maximum Gap methods evaluates the agreement between the distribution of the observed events
  with the theoretical WIMP event distribution. The fact that the coverage of the Likelihood method is closer to 90\% CL than the other one is explained by the
   incorporation of the $L(\mu_s,\mu_b)$ term in the extended likelihood function which helps at discriminating WIMP events from the background ones contained in
    a given dataset. Thus, the Likelihood method is the most robust method in the case of   a sizeable background contamination. Note that if the expected number
 of background events could be accurately estimated even with some uncertainties, it could be possible to consider a non constant prior $P(\mu_b)$ in order to improve the
  efficiency of the Likelihood method. However, as discussed in \cite{yellin1}, as the number of expected background $\mu_b$ cannot be known without important uncertainties,
   it is not possible to use the Feldman-Cousins technique \cite{feldman} and the ones presented in \cite{Cowan}. 

\section{Comparison of the exclusion limit methods}
\label{sec:Comparison}
As the statistical framework is being defined, we can compare the efficiency of the three methods. 
We focus on pure background or highly background-contaminated data, as  it is the most realistic case for upcoming directional
detectors, especially prototype experiments with low exposures. In the following, we evaluate with the three statistical methods,
  the effect of detector configurations on the exclusion limits, such as  having or not 
  sense recognition capability,   having a low angular resolution and varying the energy threshold.
  The effects of   astrophysical  uncertainties on the calculation of cross section limits are eventually evaluated.\\
Unless otherwise stated, we consider a CF$_4$ directional detector and a halo model as presented in section \ref{sec:detection}, with a total time exposure of 3 years and a 
WIMP mass taken to be $100$ GeV.c$^{-2}$. 
\begin{figure}[b]
\begin{center}
\includegraphics[scale=0.35,angle=270]{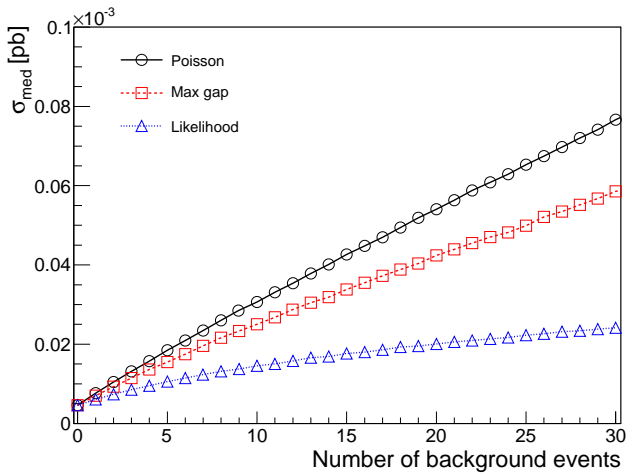}
\caption{The excluded cross section ($\sigma_{med}$) as a function of the number of background events, in the case of a pure background,
  obtained by the three different statistical methods: Poisson (black solid line), Maximum Gap (red dashed line) and Likelihood (blue dotted line).}  
\label{fig:BackgroundPureHT}
\end{center}
\end{figure}

\begin{figure}[t]
\begin{center}
\includegraphics[scale=0.35,angle=270]{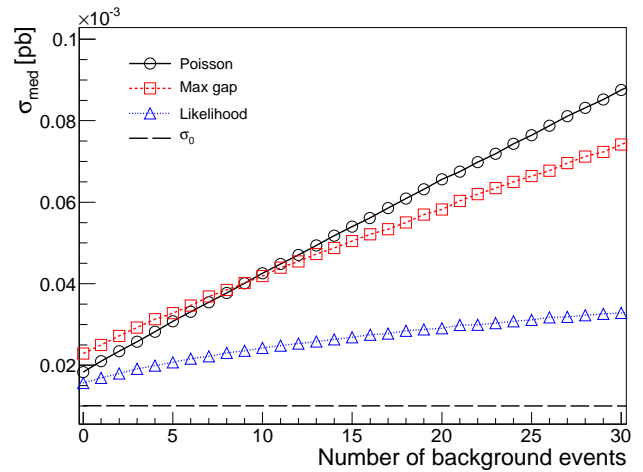}
\caption{The excluded cross section ($\sigma_{med}$) as a function of the number of background events, in the case of 5 expected WIMP events,
  obtained by the three different statistical methods: Poisson (black solid line), Maximum Gap (red dashed line) and Likelihood (blue dotted line).}  
\label{fig:5WIMP}
\end{center}
\end{figure}
\subsection{Ideal detector}
First, an "ideal detector" is considered, meaning a detector allowing a
3D reconstruction of   recoiling events 
 with sense recognition and with perfect angular definition.
  The three methods are compared on two simulated datasets : pure background (fig.~\ref{fig:BackgroundPureHT}) and background events with
5 expected WIMP events (fig.~\ref{fig:5WIMP}). The first case reflects a scenario in which the exposure is not sufficient to access 
WIMP events due to a very low WIMP-nucleon
 cross section while in the second case a few WIMP events are obtained as the considered cross section was increased.  Further an increase of the number of WIMP events is meaningless in the context of directional 
detection, as in such case a discovery could be done rather than an exclusion, as shown in \cite{billarddisco}. 
Figure \ref{fig:BackgroundPureHT} and \ref{fig:5WIMP} present  the excluded cross section ($\sigma_{med}$ as defined in sec.~\ref{sec:quescution}) obtained with the three different
methods as a function of the number of expected background events.\\ 
In the pure background case (fig.~\ref{fig:BackgroundPureHT}), the three methods start 
with  a common value at null event. Indeed, in such a case, the Maximum Gap and the Likelihood methods are identical to the Poisson one. Then, when no event is observed,
an excluded value of $\rm \mu_{exc} \simeq 2.3$ WIMP events at 90\% CL is deduced from the three methods. The excluded cross section obtained with the three methods start
 being different as 
the number of background events increases. The Likelihood method allows to obtain a much more restrictive exclusion limit than the two others, by a factor $\sim 4$ at high background
contamination (30 events). As expected, 
increasing the number of background events degrades the upper cross section limit ({\it i.e.} increases the value of $\sigma_{med}$). This holds for the three methods,
noticing that the Likelihood method is less sensitive to background contamination, as it takes full advantage on the 
knowledge of the expected WIMP and background angular distributions.\\
In the case of data populated with a few WIMP events, the excluded cross section evaluated with the Likelihood method is only 
slightly less restrictive than in the former case. The two Poisson curves (black solid lines) are exactly the same on figures \ref{fig:BackgroundPureHT} 
and \ref{fig:5WIMP}. Indeed, as the Poisson method is only dealing with the total number of events, the presence of WIMP events does not 
modify the exclusion limit. It can be noticed that below $\sim 10$ events, the Maximum Gap method  gives worse results than 
the Poisson one. As a matter of fact, in the case of pure WIMP signal the Poisson method overcomes the Maximum Gap one, see sec.~\ref{sec:Poisson}\\

\begin{figure}[t]
\begin{center}
\includegraphics[scale=0.35,angle=270]{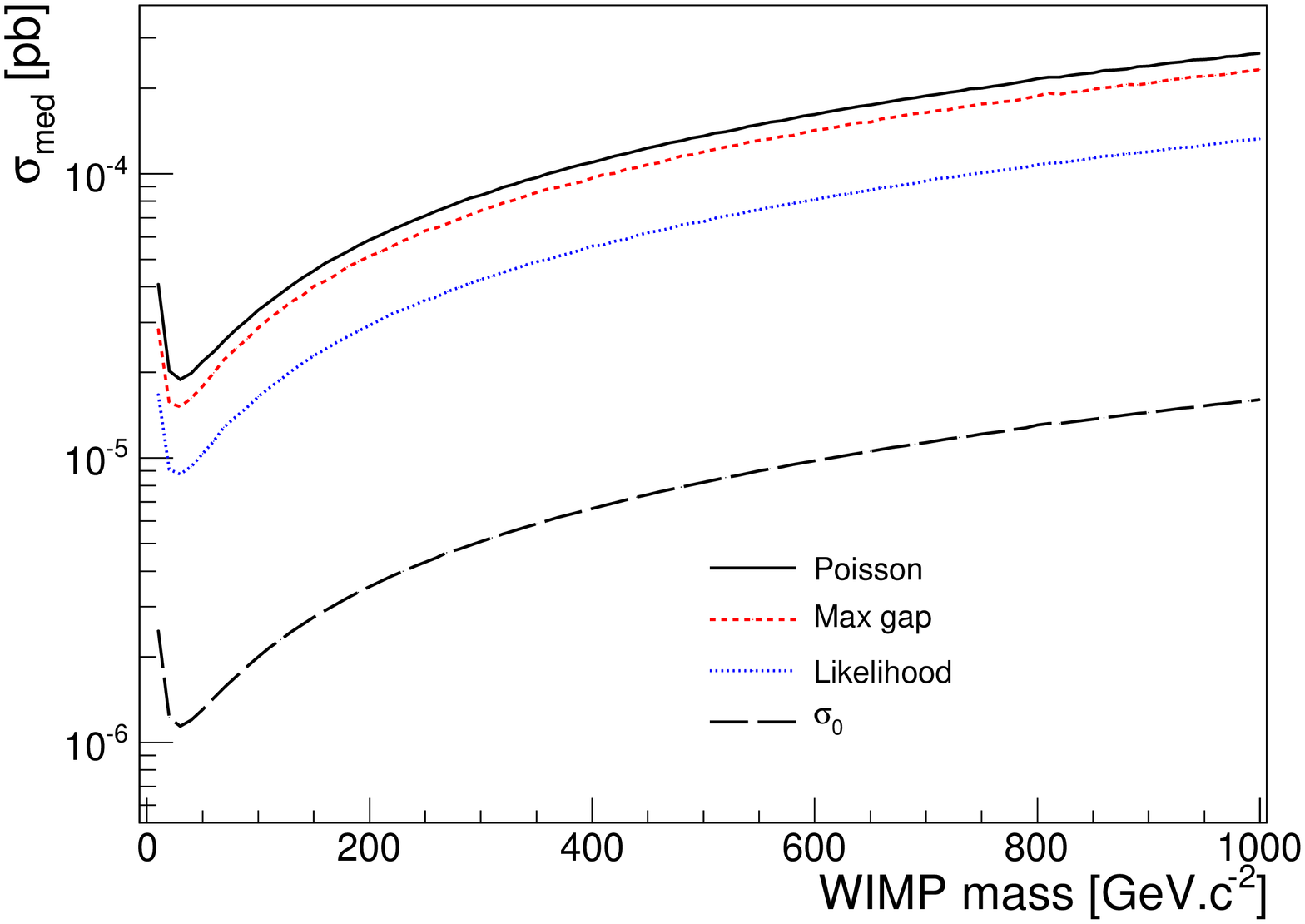}
\end{center}
\begin{center}
\vspace{2mm}
\includegraphics[scale=0.35,angle=270]{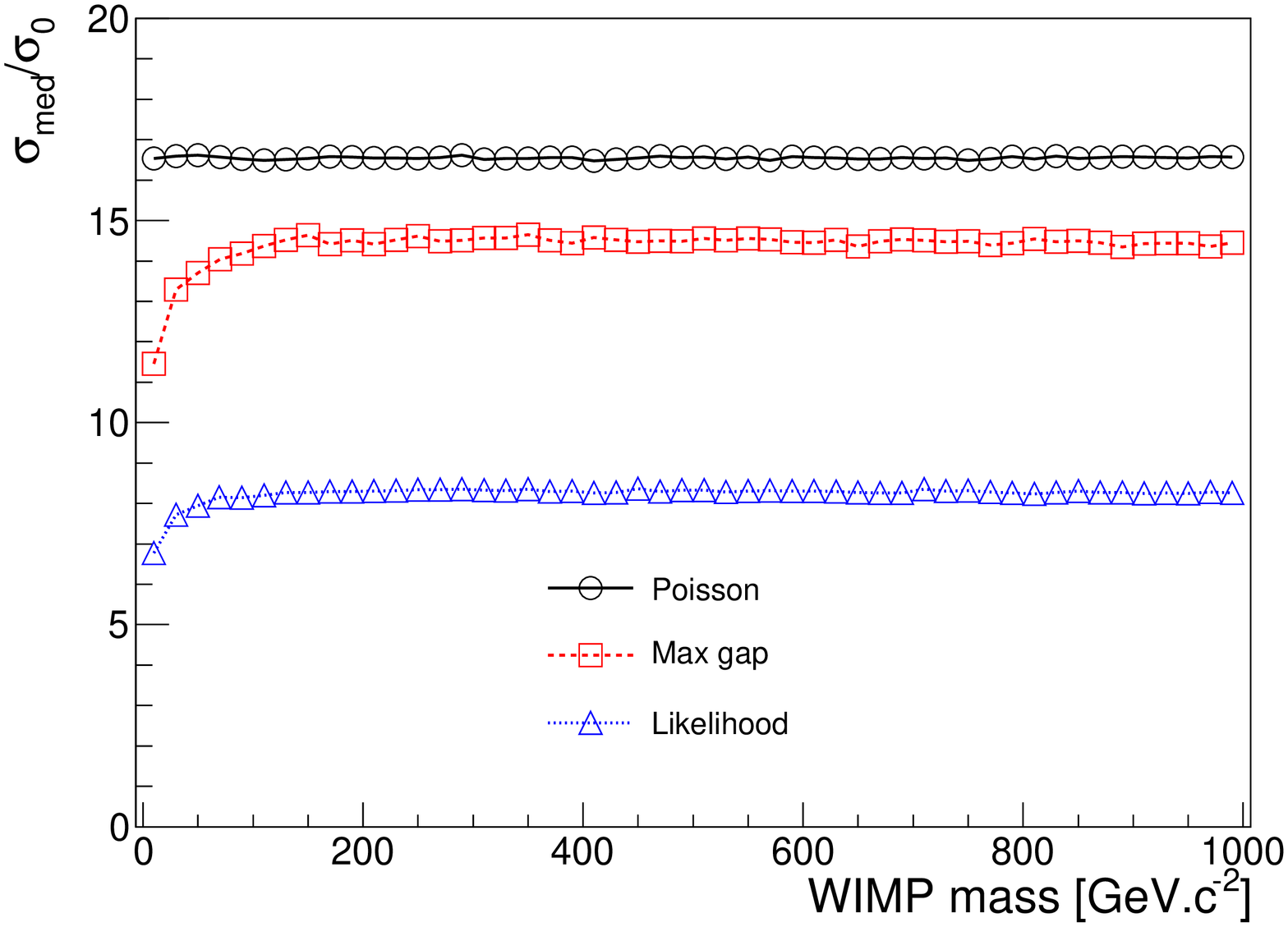}
\caption{Upper panel  :  cross section limit ($\sigma_{med}$) 
as a function of the WIMP mass for the three methods considering 10 expected background events and one WIMP
event. Lower panel :  cross section ratio ($\sigma_{med}/\sigma_0$)  as a function of the WIMP mass. This highlights the fact that the 
cross section limit is enhanced by the directionality of the signal, at light WIMP masses ($\leq 100$ GeV.c$^{-2}$), for both the Maximum 
Gap and the Likelihood method.}  
\label{fig:RapportSigmaMasse}
\end{center}
\end{figure}

As shown in \cite{billarddisco}, the directionality of the WIMP signal depends on the WIMP mass. Lighter is the WIMP, stronger is the angular anisotropy. Indeed, due to
 the finite energy range and the fact that 
 low   WIMP mass induces an energy distribution  shifted to low energy,  events above threshold ($E_{R_1}$) are the one with the most directional feature
 (eq. (\ref{RecoilEnergy})).
 The directionality evolves quickly at low masses and very slowly for masses heavier than a hundreds of GeV.c$^{-2}$.
 A study of the effect of the WIMP mass on the cross section limit has been done.
Figure \ref{fig:RapportSigmaMasse} presents on the upper panel the cross section limit ($\sigma_{med}$) 
as a function of the WIMP mass for the three methods considering 10 expected background events and one expected WIMP
event. This highlights the fact that the Likelihood method gives the best result on the
whole WIMP mass range. Moreover, we can notice that the cross section limit is more restrictive for masses lower than 
$\sim 30 \ {\rm GeV.c^{-2}}$ due to the threshold of the detector.  Indeed, at
low masses, the energy distribution is more peaked at low energy leading to a loss of expected events and then to 
an increasing cross section limit. The lower panel of figure \ref{fig:RapportSigmaMasse} presents  the evolution of
 $\sigma_{med}$ divided by $\sigma_0$ (the input value  to get 1 WIMP event). 
 It allows to show that the  cross section limit is enhanced by the directionality of the signal, at low WIMP masses ($\leq 100$ GeV.c$^{-2}$), for both the Maximum 
Gap and the Likelihood method.  Increasing the WIMP mass does not affect the result as there are little differences in angular distributions for heavy WIMPs, 
as stated above. As expected the Poisson method is not sensitive to the WIMP mass, as it handles only the total number of events 
recorded and not their angular distribution.\\

\begin{figure}[t]
\begin{center}
\includegraphics[scale=0.35,angle=270]{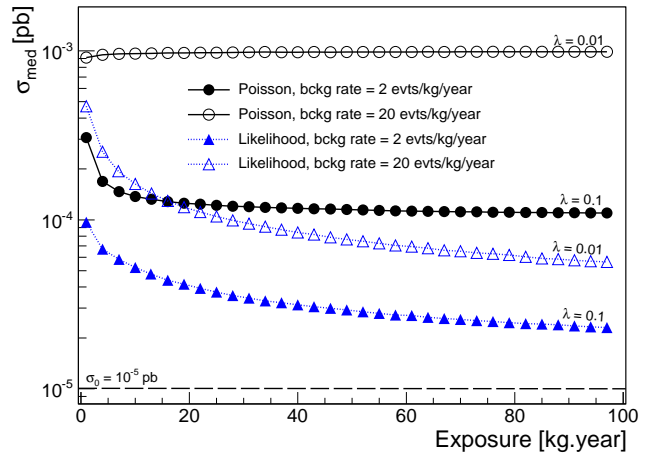}
\caption{The excluded cross section obtained with the Poisson and Likelihood methods as a function of the exposure. $\lambda$ is defined as
the WIMP fraction in the events : $\lambda=\mu_s/(\mu_s+\mu_b)$. For this study the WIMP mass is fixed $m_{\chi} = 100$ GeV.c$^{-2}$ and  
  two different background rates are considered : (2, 20) evt/kg/year. The input 
 cross section is 
$\sigma_0 = 10^{-5}$ pb giving an expected rate of WIMP events equal to 0.2 evts/kg/year.}  
\label{fig:Exposition}
\end{center}
\end{figure}

A key feature  of the proposed Likelihood method is the fact that,  contrary to the Maximum Gap method \cite{yellin1}, it can deal 
with very large number of events.
Hence, we study hereafter the evolution of $\sigma_{med}$ with the exposure, for a fixed WIMP mass 
$m_{\chi} = 100$ GeV.c$^{-2}$. Figure \ref{fig:Exposition} presents the excluded cross section obtained with the 
Poisson and Likelihood methods as a function of the exposure. Curves are labeled according to the expected value of $\lambda$, defined as
the WIMP event fraction in the dataset ($\lambda=\mu_s/(\mu_s+\mu_b)$). To illustrate, we have chosen two different background rates (2, 20) evt/kg/year  and an input 
 cross section
$\sigma_0 = 10^{-5}$ pb giving an expected rate of WIMP events equals to 0.2 evts/kg/year.
 The Poisson results (in black circles) are also represented in order to appreciate the efficiency of the Likelihood method (blue triangles).

With the Likelihood method, the cross section upper limit is still getting more restrictive, when increasing the exposure,  especially in the case of highly 
background-contaminated data. 
On the contrary, if the background fraction is too high ($\lambda \leq 0.01$), the
 cross section limit given by the Poisson method remains flat as a function of the exposure. 
 This is the proof that the Poisson method will not be able to deal with data contaminated by a large number of background events, as  expected from upcoming directional
 prototype detectors. 
Then, the difference between the Poisson and the Likelihood method is about an order of 
 magnitude on this range of exposure. But the most interesting point is the fact that only the Likelihood method will allow to set better
 exclusion limits with increasing exposure, even with critically low value of $\lambda$. 
 As expected, when increasing the rejection power of the detector, {\it i.e.} increasing the expected value of $\lambda$, 
 the two statistical    
 methods (filled triangles and filled circles) are getting more efficient.
 
\subsection{Detector without sense recognition}
\label{sec:sense}
Even though several progresses have been done~\cite{dujmic,burgos,majewski}, sense recognition (so-called "Head-Tail" (HT) effect) remains a key and challenging 
experimental issue for directional detection of Dark Matter. In particular, it should still be demonstrated that sense 
recognition may be achieved at low recoil energy where most WIMP events reside. Hence, in the following 
we investigate the effect  on exclusion limits of using a detector with  no sense recognition. 
In such case, a recoil coming from ($\cos\gamma$,$\phi$) cannot be distinguished from a recoil coming
from ($-\cos\gamma$,$\phi + \pi$).  Then, we need to use axial data characterized by a direction
   ($|\cos\gamma|$) and the directional event distribution becomes:
\begin{equation}
\frac{dN}{d|\cos\gamma|} = \frac{dN(\cos\gamma)}{d\cos\gamma} + \frac{dN(-\cos\gamma)}{d\cos\gamma}
\end{equation}

\begin{figure}[t]
\begin{center}
\includegraphics[scale=0.35,angle=270]{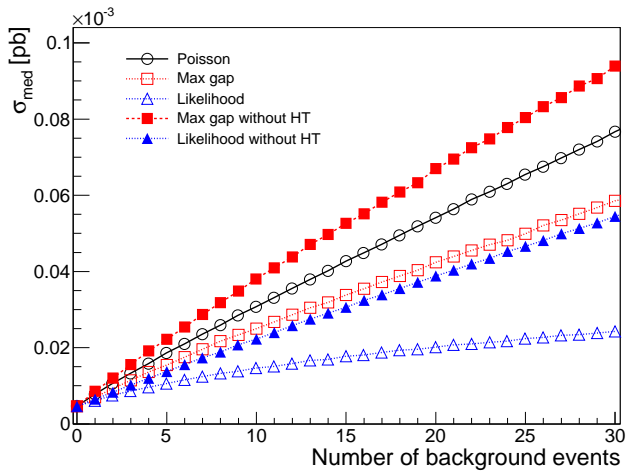}
\caption{The median upper limit cross section obtained by the three different statistical methods as a function of the number of background events in the case of pure
 background. The results without sense recognition are presented with filled markers while empty markers refer to a detector with sense recognition.}  
\label{fig:BackgroundPureSanSHT}
\end{center}
\end{figure}

We study the effect of no sense recognition in the case of pure background data and the result is presented on figure \ref{fig:BackgroundPureSanSHT}. 
To ease comparison, results with sense recognition are recalled (from figure \ref{fig:BackgroundPureHT}). Still, the Likelihood method overcomes the two others and it appears
 that the absence of 
sense recognition only mildly alter the result (a factor of three at high background contamination). 
 As expected, the Poisson method is insensitive to the absence/presence of sense recognition. Surprisingly, the Maximum Gap method becomes less competitive than the
 Poisson method in the case of no sense recognition. This is due to the fact that in this case, the expected WIMP distribution becomes less anisotropic and then it is getting closer
  to the 
 expected background event distribution (isotropic). Indeed, as it has been shown in \cite{yellin1,henderson}, in the case where the difference between the 
 background and the WIMP event distribution is very small, the Poisson method sets more competitive limits than the Maximum Gap one.\\
Taken at face value, this result suggests that sense recognition may not be so important for directional detection when trying to set exclusion limits. The difference between 100\% sense recognition on the
whole recoil energy range, which is obviously unrealistic, and no sense recognition is only minor (less than a factor of three at high background contamination). The worst case is indeed a partial
sense recognition strongly depending on the recoil energy. In this case, we suggest not to consider this information to set exclusion limits. However, sense recognition remains a key issue
worth investigating,  for WIMP discovery which is the ultimate goal of directional detection.


\subsection{Detector with finite angular resolution}
Previous studies have been done in the case of a detector with a perfect angular resolution. However, realistic data of upcoming directional detectors
should suffer from finite angular resolution. This is an intrinsic limitation of this detection strategy. Even if simulations
 show that straight line tracks may be 3D reconstructed with a rather small angular dispersion \cite{grignon}, realistic tracks in low
 pressure gaseous detectors would encounter a rather large "straggling" effect (angular dispersion). The lower is the recoil energy, the larger is the angular
 straggling. Hence, in the following 
we investigate the effect  on exclusion limits of using a detector with  a finite (realistic) angular resolution. Having a finite angular
resolution means that a recoil initially coming from the direction  $\hat{r}(\theta, \phi)$ is 
reconstructed as a recoil ${\hat{r}}^{\, \prime}(\theta^\prime, \phi^\prime)$. Then, the angular deviation between 
the initial recoil and the reconstructed one is given by:
\begin{equation}
\Theta = \cos^{-1}(\hat{r} \ . \ {\hat{r}}^{\, \prime})
\end{equation}
The ${\hat{r}}^{\, \prime}$ direction is characterized in the $\hat{r}$ frame 
by the angles $\Theta$ and $\Phi$ which are, for each event, randomly generated using the following probability 
density function $f(\Theta,\Phi)$:
\begin{equation}
f(\Theta,\Phi) \propto \sin\Theta\exp\left( -\frac{\Theta^2}{ 2\sigma^2_{\Theta}}    \right)
\begin{array}{rl}
& \ 0\leq\Theta\leq \pi \\
& \ 0 \leq\Phi\leq 2\pi 
\end{array}
\end{equation}
In order to compute the theoretical angular distribution considering this finite angular resolution, we have done 
the convolution
 product of the initial angular spectra $dN/d\Omega$ with a smoothing gaussian function of 
 width $\sigma_\Theta$ \cite{copi2}.\\
 
\begin{figure}[t]
\begin{center}
\includegraphics[scale=0.35,angle=270]{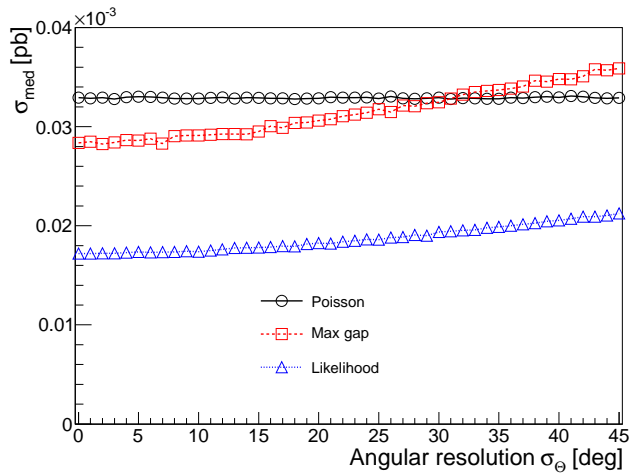}
\caption{The median upper limit cross section $\sigma_{med}$ obtained with the three methods
 as a function of the angular resolution. The study is done in the case of 10 expected 
background events and one expected WIMP
event. The detector configuration is the following :  sense recognition and energy threshold of 5 keV.}  
\label{fig:RapportSigmaResolution}
\end{center}
\end{figure}

Figure \ref{fig:RapportSigmaResolution} presents the cross section limit as a function 
of the angular resolution $\sigma_{\Theta}$ in the case of 10 expected background events and one expected WIMP
event. 
It can be noticed that the Maximum Gap and the likelihood methods are only slightly dependent on the angular 
resolution of the detector. The  deviation for the two methods is of the order of 30\ \%
from $\sigma_{\Theta} = 0^{\circ}$  to $\sigma_{\Theta} = 45^{\circ}$. Hence, as far as exclusion limits 
 are concerned, the effect of angular resolution is relatively small but has to be correctly evaluated to set coherent upper limits.
  This outlines the need for detector commissioning, e.g. by using a neutron field \cite{amokrane}.\\
 As for the sense recognition study (sec. \ref{sec:sense}) and for the same reasons, the Maximum Gap method is getting less competitive than the Poisson method beyond
 $\sigma_{\Theta} \sim 30^{\circ} $. Indeed, the expected WIMP distribution and background are getting closer and then, as previously explained,
  the Poisson method is more competitive than the Maximum Gap one.

\subsection{Effect of energy threshold}
\begin{figure}[t]
\begin{center}
\includegraphics[scale=0.35,angle=270]{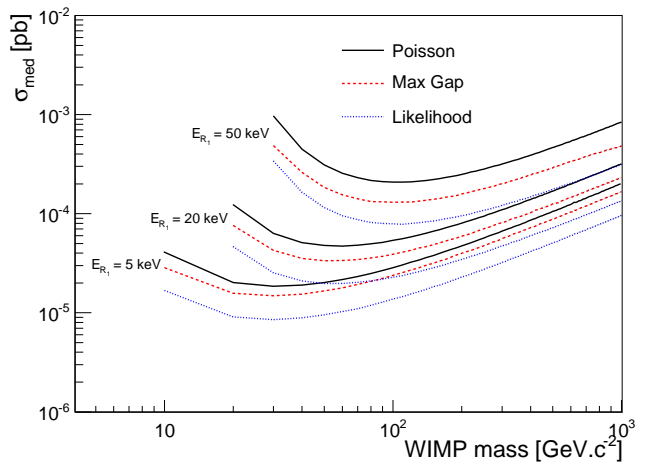}
\caption{Exclusion limits, WIMP-nucleon cross section as a function of the WIMP mass 
with the three methods and for three values of the energy threshold. The study is done in the case of 10 expected 
background events and one expected WIMP
event. The upper bound of the energy range $\rm E_{R_2}$ is chosen as 200 keV.  The detector configuration is the following : perfect angular
resolution and sense recognition.}  
\label{fig:threshold}
\end{center}
\end{figure}

As for direction-insensitive direct detection, the energy threshold plays a key role for directional detection. It depends on the target and the quenching factor.
 It is defined as the
minimal recoil energy for which both the energy and the 3D track are well reconstructed. Reducing the energy threshold leads to a higher
expected number of WIMP events and hence a more restrictive exclusion limit.  Figure \ref{fig:threshold} presents the exclusion limits obtained, 
evaluated by the three statistical methods and for three different values of the energy threshold $\rm E_{R_1}$,
in the case of 10 expected background events and one expected WIMP
event. As the energy threshold $\rm E_{R_1}$ is taken between 5 and 50 keV, the upper bound of the energy range $\rm E_{R_2}$ is chosen as 200 keV for 
this study. It can be noticed that the Likelihood
method gives the best limit on the whole mass range and for the three threshold values. 
It is worth emphasizing that the effect of the energy threshold is very important even in the case of directional detection. Going from 
a 5 keV
to a 50 keV energy threshold leads to a loss of about one order of magnitude in exclusion limits. The situation is even worse at low WIMP masses. This outlines the
fact that, as far as exclusion limits are concerned, lowering the energy threshold remains the major experimental issue 
for upcoming directional detector projects.
  
\subsection{Impact on limits from astrophysical uncertainties}
\label{sec:vo}
The effect of astrophysical uncertainties on exclusion limits obtained with direction-insensitive direct detection has been studied 
in details in \cite{green.moriond.ew10,mccabe}. In the
following, we investigate the effect on directional detection.\\
Within the framework of a spherical isothermal halo model, two astrophysical parameters play a key role : 
the  local dark matter density $\rho_0$ and the local WIMP velocity dispersion $v_0$.\\ 
The first one is usually quoted within the range  $\rho_0 \sim 0.2 - 0.8$ GeV.c$^{-2}$.cm$^{-3}$ and the value 
$\rm 0.3 \  GeV.c^{-2}$ is used as a "standard" value for the sake of comparison of    various direct 
detector results  \cite{pdg}. Recently, a value of the local Dark Matter density has been derived within the framework of a galaxy-model independent
method \cite{salucci}.  The resulting local Dark Matter is $\rho_0 = 0.43  \pm 0.11 \pm 0.10 \ {\rm  GeV/c^2/cm^3}$. As shown in
\cite{billarddisco}, directional detection offers the possibility to constrain the WIMP-nucleon cross section, 
which can be relaxed into a
constraint on $\rho_0 \times \sigma$ as a function of the WIMP mass. Then, the measured value of $\rho_0$ \cite{salucci},
 together with its uncertainty, shall be accounted for when presenting discovery regions. However, when setting limits,  the
 value of $\rho_0$ does not change  the shape of the exclusion limit, only its magnitude and with a negligible effect, owing to the orders of
 magnitude involved. In the following the so-called standard value is used.\\

\begin{figure}[h]
\begin{center}
\includegraphics[scale=0.35,angle=270]{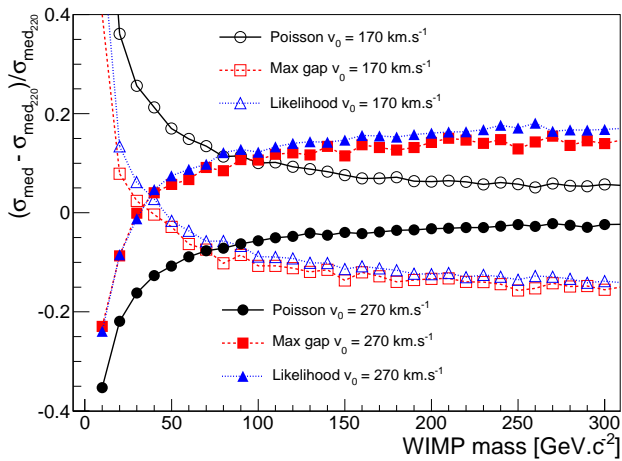}
\caption{Deviation of the median upper limit cross section ($\sigma_{med}$) from the value $\sigma_{med_{220}}$ 
obtained with the standard value  
$v_0=220 \ {\rm km.s^{-1}}$, as a function of the WIMP mass. The study is done, for the three methods, with two extreme cases :   170 km.s$^{-1}$  and 270 km.s$^{-1}$.}  
\label{fig:SigmaRapportMasseV0}
\end{center}
\end{figure}

The second astrophysical parameter to be carefully handled is the WIMP velocity dispersion $\sigma_v$ related to the asymptotic circular velocity $v_0$,
 for which the "standard" value is 
$220 \ {\rm km.s^{-1}}$ \cite{pdg}. As outlined in 
\cite{green.moriond.ew10}, recent determinations of its value span on a wide range and the impact for directionality 
has been studied, in the context of isotropy rejection~\cite{copi2}. In the following, we investigate the effect of its value   
on directional exclusion limits by studying  two extreme cases ($170$ and $270 \ {\rm km.s^{-1}}$).\\ 
Figure \ref{fig:SigmaRapportMasseV0} presents the deviation of the median upper limit cross section ($\sigma_{med}$) from the 
value $\sigma_{med_{220}}$ obtained with the standard value  $v_0=220 \ {\rm km.s^{-1}}$ as a function of the WIMP mass. 
The study is done with a dataset of 10 background events and 1 WIMP event.  
For convenience, the Poisson case, represented by the black solid lines, is described first. As stated above, the Poisson exclusion limit 
is only sensitive to the total number of events and not to the directionality of the event angular distribution. 
When comparing low ($170 \ {\rm km.s^{-1}}$, open circle) and high ($270 \ {\rm km.s^{-1}}$, filled circle) values of $v_0$, three effects on the excluded cross section
 can be noticed. The first one is linked to the energy threshold. For  
a low value of $v_0$, the theoretical WIMP energy distribution is more peaked at low energy, leading to less events above 
the energy threshold. Hence,  the cross section limit is higher in the case of  $v_0=170 \ {\rm km.s^{-1}}$ than for $v_0=270 \ {\rm km.s^{-1}}$.
A second effect is due to the WIMP mass. Indeed, the difference between low and high  values of $v_0$ is increased for low WIMP masses. 
Again, this is due a shift of a fraction of WIMP events below the threshold at low WIMP masses.\\
When comparing with  Maximum Gap and the likelihood methods, a third effect  is introduced due to the
 directionality of the WIMP signal. The same general tendency is observed,  
 but above  a given WIMP mass ($\sim 40  \ {\rm GeV/c^2}$), the low $v_0$ exclusion limit becomes more restrictive. Indeed, a lower value of $v_0$ will 
 induce a more directional WIMP event distribution as it is linked to the spread of the WIMP flux, see eq. \ref{eq:HaloModel}. Then, the difference
 between the background and the WIMP signal is enhanced, leading to a better exclusion. To summarize, when the WIMP mass is 
 sufficiently high, the effect of the directionality at low $v_0$ overcomes the induced loss of events previously described.\\
 As a conclusion of this study, we can notice that the impact of uncertainties on the value of $v_0$ can lead to a deviation of the excluded cross section
  from 50\% at low WIMP mass to 20\% at large WIMP mass.\\
  
As stated above,  for this first study of an optimal exclusion method for directional detection, the standard halo model 
(spherical isothermal) has 
been considered.  This is indeed the reference model to compare Dark Matter experiments and associated exclusion methods.
Anisotropy, triaxiality,  and clumping in the WIMP velocity distribution would   have an effect 
on  directional detection signal. As far as directional exclusion limits are concerned, this will 
mildly affect the result presented in this paper. Indeed, only the theoretical WIMP
distribution used for the directional likelihood method would be changed, though  remaining directional (anisotropic) and then 
different from the background. However, the effect of non standard halo models (triaxial, with stochastic features or streams) will be addressed in 
       a dedicated forthcoming paper \cite{next}.


   \begin{figure}[t!]
\begin{center}
\includegraphics[scale=0.48,angle=0]{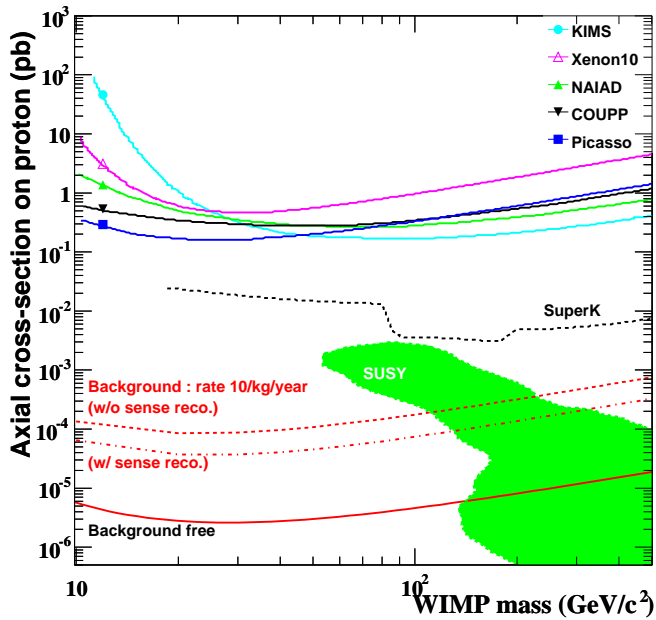}
\caption{Spin dependent cross section on proton (pb) as a function of the WIMP mass ($\rm GeV/c^2$). Results are presented
in the case of pure-proton approximation \cite{tovey}. 
The theoretical region, obtained within the framework of the constrained minimal 
 supersymmetric model, is taken from \cite{superbayes}. Constraints from collider data and relic abundance are accounted for.
Exclusion limits from direct and indirect Dark Matter searches are displayed (see text for details). 
The projected exclusion limit of a forthcoming directional detector (30 kg.year) is presented in three cases : 
background-free (solid line), with a background rate of 10 events/kg/year with sense recognition (dot-dashed line) and without sense recognition (dashed line).} 
\label{fig:ResultMimactot}
\end{center}
\end{figure}

\section{Prospecting exclusion limits with directional detection}
\label{prospects}
To end-up this study, we present projected exclusion limits for a forthcoming directional detector proposed by the MIMAC collaboration \cite{MIMAC}. We consider 
a  10 kg $\rm CF_4$ detector 
operated during  $\sim 3$ years,  allowing 3D 
track reconstruction, with a $10^\circ$ angular resolution and a  recoil energy range 5-50 keV.
In the following, a very low WIMP-nucleon cross section is considered, such as no WIMP event 
is expected during the acquisition time of the experiment. Note that a detector configuration has been chosen as an illustration, but the effect of the 
energy range, angular resolution has been shown previously.\\
Figure \ref{fig:ResultMimactot} presents the spin dependent  cross section on proton (pb) as a function of the WIMP mass ($\rm GeV/c^2$). Results are presented
in the case of pure-proton approximation \cite{tovey}. The proton spin content has been chosen as $\langle S_p \rangle=0.441$ 
\cite{pacheco}. Exclusion limits from direct detection searches are presented :   COUPP~\cite{coupp},  
KIMS~\cite{kims}, NAIAD~\cite{naiad}, Picasso~\cite{picasso} and  Xenon10~\cite{xenon10}.
Exclusion limits obtained with neutrino telescopes (Super K~\cite{superk}) are also displayed.  
The theoretical region, obtained within the framework of the constrained minimal supersymmetric model, 
is taken from \cite{superbayes}. Constraints from collider data and relic abundance $\Omega_\chi h^2$,
 as measured with WMAP 5-year data \cite{wmap5}, are accounted for.\\ 
 Several cases are considered on figure \ref{fig:ResultMimactot} : 
\begin{itemize}
\item the background-free measurement (dashed curve). It stands as the ultimate exclusion limit for a directional detector with such an
exposure.
\item with a background rate of 
10 events/kg/year, which is large compared to current background measurements \cite{edelweiss-armengaud}.
We consider the case of a directional detector with sense-recognition (dot-dashed line). The loss due to the
presence of background is of the order of one order of magnitude, noticing that the use of the proposed Likelihood method allows the result
to remain very satisfactory. 
\item with a background rate of 
10 events/kg/year but for a detector without sense-recognition (dashed line). As stated above, as far as exclusion limits are 
concerned, the effect of not having sense-recognition is small. 
\end{itemize}
It highlights the exclusion power of a  rather light directional 
 detector (10 kg), in the  case that the discovery analysis \cite{billarddisco} failed at recognizing the galactic 
 origin of the signal,  for instance if the axial nucleon-WIMP cross section is very low. Even with highly contaminated data or
 without sense recognition directional detection with a 10 kg CF$_4$ detector would allow to set very constraining exclusion limits, about 3 orders
 of magnitude better than the one imposed by existing direct detectors.

\section{Conclusion}
\label{sec:Conclusions}
The conclusion of this study   is threefold.\\ 
First, a new likelihood method to derive exclusion limits 
has been proposed for the directional search of galactic WIMPs. Only the angular part of the spectrum is taken into account (in a given energy
range),
 arguing that the energy part of the WIMP spectrum is featureless and may even be mimic by the background one. 
 On the same realistic set of directional data, the method has been shown to be more competitive than existing ones, or derived from
 existing ones.\\
Second, in the spirit of directional detector opimization, a study of the effect of detector configurations on the exclusion limits has
been done. Comparing the results from the three methods shows that the proposed exclusion Likelihood method gives the best results in all
cases. Moreover, the effect of not having sense recognition capability as well as having a low angular resolution has been shown to have
 little effect on exclusion limits. On the contrary, the effect of the energy threshold has been shown to largely influence the 
 exclusion limits. This highlights the fact that the energy threshold is definitely the major experimental challenge for upcoming directional
 detectors, while no sense recognition and low angular resolution could be sufficient as far as exclusion limits are concerned, which might be the case for low exposure data.
  Note that
 this conclusion does not hold for a discovery strategy \cite{billarddisco}, {\it e.g.} if the unknown axial WIMP-nucleon cross section lies  
 in  the   $10^{-3} \ {\rm pb}$ region. In this case, these 
 detection issues remain of major   interest, worth pursuing
 experimental efforts.\\
 Third,  the exclusion power of a   light directional 
 detector (10 kg) has been shown to be competitive with respect to the existing exclusion limits. Even with highly contaminated data or
 without sense recognition it allows to set exclusion limits about 3 orders
 of magnitude better than the one imposed by existing direct detectors. 
 

\end{document}